\documentclass{article} 
\usepackage{iclr2027_conference,times}

\usepackage{amsmath,amsfonts,bm}

\def\eqref#1{equation~\ref{#1}}

\def\1{\bm{1}}

\DeclareMathAlphabet{\mathsfit}{\encodingdefault}{\sfdefault}{m}{sl}
\SetMathAlphabet{\mathsfit}{bold}{\encodingdefault}{\sfdefault}{bx}{n}

\usepackage{capt-of}
\usepackage{hyperref}
\usepackage{url}
\usepackage{booktabs}
\usepackage{amsmath}
\usepackage{graphicx}
\usepackage{tabularx}
\usepackage{makecell}
\usepackage{pifont}
\usepackage{capt-of}
\usepackage{wrapfig}
\usepackage{placeins}
\usepackage{float}
\usepackage[table]{xcolor}
\usepackage{colortbl}   
\newcommand{\cmark}{\textcolor{teal!70!black}{\ding{51}}}
\newcommand{\xmark}{\textcolor{magenta!70!black}{\ding{55}}}
\newcommand{\pmark}{\textcolor{orange!80!black}{\ding{108}}}
\newcommand{\jtype}[1]{\textit{#1}}
\hypersetup{
    colorlinks=true,
    linkcolor=black,
    citecolor=black,
    urlcolor=blue
}
\title{
\textbf{MuLA-Bench}: A Multilingual Long-Form Audio Understanding Benchmark via Multi-Tier Auditing
}

\author{
\textbf{Zeyu Yang}\textsuperscript{1,2},
\textbf{Xinyu Zhang}\textsuperscript{2}\thanks{Corresponding authors.},
\textbf{Zibo Bi}\textsuperscript{2,3},
\textbf{Pei Zhang}\textsuperscript{2},
\textbf{Xize Cheng}\textsuperscript{2},
\textbf{Jin Xu}\textsuperscript{2},
\\
\textbf{Baosong Yang}\textsuperscript{2},
\textbf{Satoshi Nakamura}\textsuperscript{1}\footnotemark[1]
\\[2pt]
\textsuperscript{1}The Chinese University of Hong Kong, Shenzhen
\quad
\textsuperscript{2}Alibaba Token Hub, Alibaba Group
\\
\textsuperscript{3}Xi'an Jiaotong University
\\[2pt]
\texttt{zeyuyang1@link.cuhk.edu.cn}
\quad
\texttt{zxy440266@alibaba-inc.com}
\\
\texttt{snakamura@cuhk.edu.cn}
}
\iclrfinalcopy 
\begin{document}

\maketitle
\fancyhead{}
\renewcommand{\headrulewidth}{0pt}
\begin{abstract}
Long-form audio performance is often summarized by context length and aggregate
accuracy, obscuring how language, evidence, and task jointly shape difficulty.
We introduce \textbf{MuLA-Bench}: 5,038 open-ended questions over 1,769
in-the-wild recordings totaling 1,377.9 hours, covering 16 languages and eight
domains. A balanced Language$\times$Domain semantic track supports controlled
comparisons, while a complementary acoustic track preserves naturally occurring
non-speech evidence. Evidence-grounded generation, shortcut checks, and
language-expert review provide auditable questions without translating a shared
source set or injecting target sounds. We evaluate ten audio-language models
and conduct pooled diagnostics on a fixed eight-model cohort. Language rankings
change across domains and tasks; acoustic--semantic performance gaps vary with
the requested operation; and temporal errors can persist after the correct
event is identified. Long-range retrieval is comparatively strong, while
precise clock alignment and factual grounding of natural acoustic events remain
fragile. MuLA-Bench thus exposes conditional failure patterns that a single
long-context score does not capture. MuLA-Bench is available at
\url{https://github.com/QwenLM/Omnilingua-Bench/tree/main/MuLA-Bench}.
\end{abstract}

\section{Introduction}
Large audio-language models (LALMs) increasingly process entire lectures,
interviews, and other long recordings. Recent benchmarks have expanded from
broad perception to reasoning and hour-scale comprehension
\citep{yang2024airbench,wang2025audiobench,sakshi2025mmau,ma2025mmar,
ahia2025blab,huang2026audiospan}. Yet a longer input window does not by itself
establish reliable understanding: the same model may retrieve a distant fact
while failing to identify a nearby sound or locate an understood event in time.
Evaluation must distinguish these operations across the languages and content
encountered in real recordings.

Existing designs offer different forms of control. Translated multilingual
variants and composed or injected events simplify part of the natural
variation; intact recordings preserve that variation but make its components
harder to disentangle \citep{gao2025adubench,iyer2026scenebench,
ye2026voicegiraffe,huang2026audiospan}. The benchmark-design question is how to
control question allocation while preserving native linguistic content and
naturally occurring acoustic evidence.

\begin{figure}[t]
\centering
\includegraphics[width=.94\textwidth]{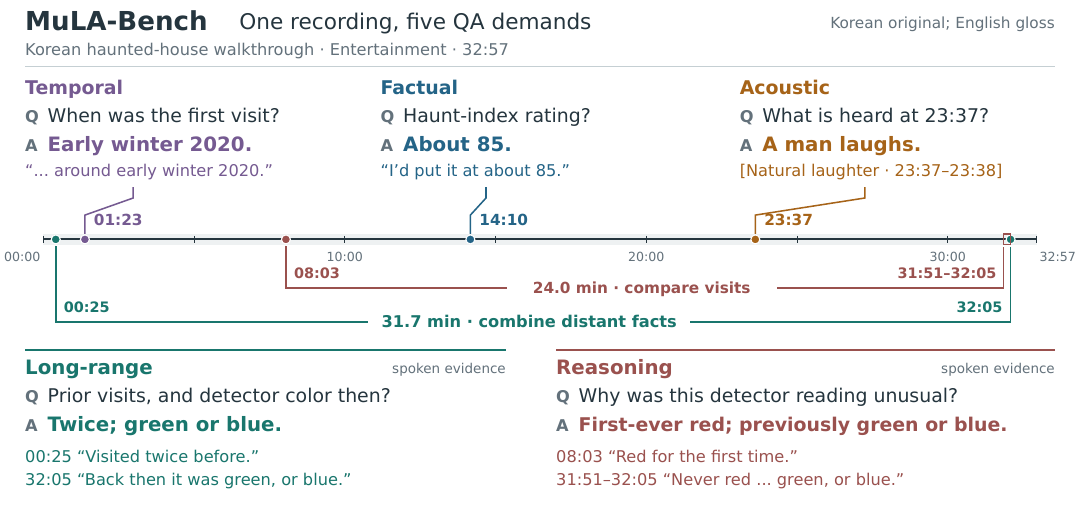}
\caption{MuLA-Bench crosses languages and domains while grounding open-ended
questions in spoken content and naturally occurring sounds from the same
source collection. Evaluated models receive audio and questions only.}
\label{fig:benchmark_overview}
\end{figure}

\textbf{MuLA-Bench} addresses this question with 5,038 questions across 16
languages and eight domains (Figure~\ref{fig:benchmark_overview}). Its semantic
track contributes exactly 30 questions per Language$\times$Domain cell, with
the same task allocation. Its acoustic track follows verified event
availability rather than imposing artificial balance. Both tracks use localized
evidence, independent question and answer calls, automatic checks, and final
language-expert review.

The resulting evaluation reveals that language difficulty changes with domain
and task, acoustic evidence carries different performance gaps across
operations, and event understanding can dissociate from physical-clock
alignment. These findings motivate measuring \emph{conditional competence}
rather than reducing long-audio understanding to one language ranking or one
length penalty. Our contributions are (i) a native multilingual benchmark with controlled
semantic allocation and natural acoustic evidence; (ii) evidence-grounded
construction with automatic and expert auditing; and (iii) ten-model evaluation
exposing conditional language rankings, operation-dependent evidence gaps,
and temporal grounding failures.

\section{Related Work}
Audio benchmarks increasingly cover perception, reasoning, and mixed
speech--sound understanding \citep{wang2025audiobench,sakshi2025mmau,ma2025mmar}.
LongSpeech, ChronosAudio, and BLAB extend evaluation to longer contexts
\citep{yang2026longspeech,luo2026chronosaudio,ahia2025blab}; VoiceGiraffe and
AudioSpan evaluate real-world English and Chinese long-form recordings
\citep{ye2026voicegiraffe,huang2026audiospan}. MuLA-Bench emphasizes how
performance changes jointly with language, domain, and evidence source.

Multilinguality and acoustic grounding introduce further design choices.
ADU-Bench translates a shared English source into additional languages, while
SCENEBench primarily uses composed scenes \citep{gao2025adubench,
iyer2026scenebench}. AudioSpan combines Native QA with injected Anchor QA;
VoiceGiraffe includes naturally interleaved speech and sound
\citep{huang2026audiospan,ye2026voicegiraffe}. As summarized in
Table~\ref{tab:benchmark_comparison}, MuLA-Bench combines broad native-language
coverage with localized natural evidence and controlled semantic question
allocation over Language$\times$Domain cells. This organization supports
conditional comparisons without replacing the original recording content.
\begin{table}[t]
\centering
\caption{
\textbf{MuLA-Bench} vs. representative long-form and multilingual audio benchmarks.
}
\label{tab:benchmark_comparison}

\small
\setlength{\tabcolsep}{4pt}
\renewcommand{\arraystretch}{1.10}

\begin{tabular*}{\textwidth}{
@{\extracolsep{\fill}}
lcccccccc
@{}
}
\toprule
&
\multicolumn{2}{c}{\textbf{Scale}}
&
\multicolumn{2}{c}{\textbf{Source}}
&
\multicolumn{2}{c}{\textbf{Evidence}}
&
\multicolumn{2}{c}{\textbf{Evaluation}}
\\
\cmidrule(lr){2-3}
\cmidrule(lr){4-5}
\cmidrule(lr){6-7}
\cmidrule(lr){8-9}
\textbf{Benchmark}
& \textbf{Dur.}
& \textbf{Lang.}
& \textbf{Native}
& \textbf{Wild}
& \textbf{Ground.}
& \makecell{\textbf{Natural} \textbf{Aco.}}
& \textbf{Open}
& \textbf{L$\times$D}
\\
\midrule
\rowcolor{gray!8}
ADU-Bench {\scriptsize\citep{gao2025adubench}}
& Dialogue & 9 & Trans. & \pmark & QA & \pmark & \cmark & \xmark \\
SCENEBench {\scriptsize\citep{iyer2026scenebench}}
& Short & 4+ & Mixed & \xmark & Task & \pmark & \pmark & \xmark \\
\rowcolor{gray!8}
ChronosAudio {\scriptsize\citep{luo2026chronosaudio}}
& $\leq$20m & 1 & -- & \xmark & Task & \xmark & \pmark & \xmark \\
BLAB {\scriptsize\citep{ahia2025blab}}
& $\sim$51m & 1 & -- & \cmark & QA & \cmark & \pmark & \xmark \\
\rowcolor{gray!8}
VoiceGiraffe {\scriptsize\citep{ye2026voicegiraffe}}
& $\sim$55m & 2 & \cmark & \cmark & Explicit & \cmark & \xmark & \xmark \\
AudioSpan {\scriptsize\citep{huang2026audiospan}}
& 10m--2h$^{+}$ & 2 & \cmark & \cmark & Explicit & \pmark & \pmark & \xmark \\
\midrule
\rowcolor{gray!12}
\textbf{MuLA-Bench}
& \textbf{20m--3h}
& \textbf{16}
& \cmark
& \cmark
& \textbf{Dual}
& \cmark
& \cmark
& \cmark
\\
\bottomrule
\end{tabular*}

\vspace{2pt}
\begin{minipage}{\textwidth}
\footnotesize
\textit{Ground.}: evidence grounding;
\textit{Natural Aco.}: naturally occurring acoustic evidence.
\cmark~fully supported;
\pmark~partially/mixed;
\xmark~not supported.
\textit{Dual} denotes localized evidence from both speech and non-speech sound.
\end{minipage}
\end{table}

\section{Benchmark Design}
\label{sec:benchmark}
\begin{figure}[t]
    \centering
    \includegraphics[width=\textwidth]{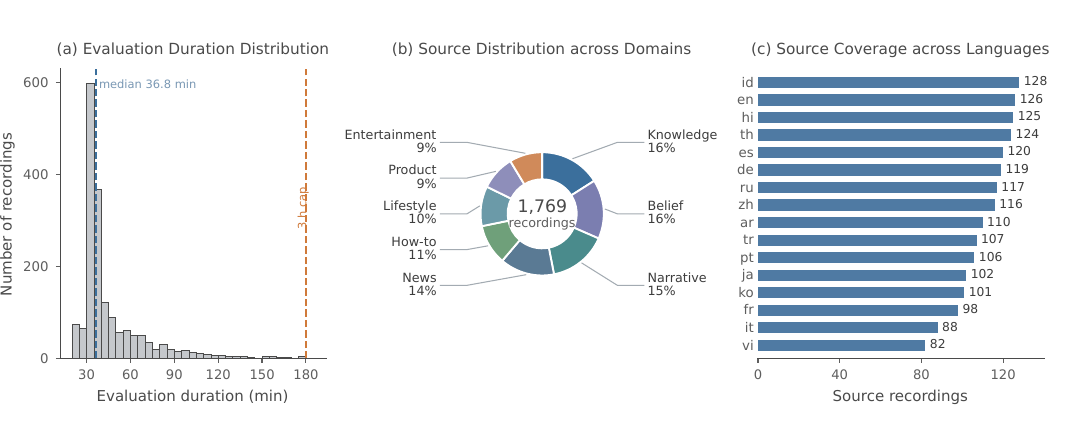}
    \caption{
    Dataset composition of MuLA-Bench.
    (a) Distribution of evaluation-audio duration, with each recording capped
    at three hours.
    (b) Distribution of source recordings across the eight content domains.
    (c) Source-recording coverage across the 16 target languages.
    }
    \label{fig:dataset_composition}
\end{figure}

\paragraph{Evaluation space.}
MuLA-Bench contains 5,038 questions over 1,769 recordings totaling 1,377.9 hours.
The median source duration is 36.8 minutes. Sixteen languages and eight domains
form 128 Language$\times$Domain cells. The domains cover knowledge, how-to,
narrative, entertainment, lifestyle, news, product, and belief. Language is a
recording-level label supported by the catalog and matching-language subtitles;
confirmed dubbing and simultaneous overlays are filtered. Native-language here
means original source content, rather than a translated benchmark variant.
Figure~\ref{fig:dataset_composition} summarizes source duration and coverage;
Appendix~\ref{app:specification} gives the full taxonomy and counts.

\paragraph{Controlled allocation, natural evidence.}
The semantic track contains 3,840 questions: exactly 30 per cell, allocated to
eight factual, eight reasoning, seven temporal, and seven long-range questions.
This balance reduces task-mixture confounding in language--domain comparisons.
The 1,198 acoustic questions instead follow verified natural event availability.
Sparse acoustic cells are not filled with translated questions or inserted
sounds. Semantic evidence concerns spoken facts and relations; acoustic evidence
concerns non-speech or paralinguistic events such as laughter, applause, and
music. Both tracks retain localized evidence; transcripts and evidence
annotations are withheld from evaluated models.

The controlled unit is the question cell, not the number or identity of source
recordings. Every semantic cell has the same task composition, so a domain's
mean gives equal weight to the 16 languages and a language's mean gives equal
weight to the eight domains. This design supports the additive-residual
analysis in Section~\ref{sec:conditional_language_rankings}. It does not make
native content across languages a parallel corpus: topics, speakers, and
recording conditions remain naturally variable. Accordingly, our language
comparisons characterize performance on this evaluation space.

Acoustic evidence uses a different sampling principle because a desired event
may not occur in a particular source. Preserving verified event availability
avoids manufacturing a balanced sound distribution. Of the acoustic items,
867 are factual, 122 reasoning, 112 temporal, and 97 long-range. We report these
counts when comparing tracks because the aggregate acoustic score is dominated
by factual items; a single semantic--acoustic difference therefore cannot
represent all four requested operations.

\paragraph{Requested operations.}
\textbf{Factual/completeness} questions request explicitly recoverable facts,
including single facts, exhaustive lists, and multiple required slots.
\textbf{Reasoning} requires inference over the recording rather than generic
world knowledge. \textbf{Temporal localization} includes event ordering and
physical time-point localization. \textbf{Long-range retrieval} uses late or
widely separated evidence: semantic items require evidence beginning at least
10 minutes into the recording or, for multiple evidence points, a span of at
least 10 minutes. Multiple points must be topically coherent. Acoustic
long-range items use separately verified event evidence; they are not
retroactively subjected to the semantic packing rule. Formal definitions and
counts are in Appendix~\ref{app:specification}.

\paragraph{Evidence for verification and diagnosis.}
Localized evidence ties answers to source content, supports expert replay, and
permits direct analysis of position and separation. A single late fact can
qualify as long-range without requiring multi-hop reasoning. Event ordering
and metric time-point localization remain distinct operations, so retrieval
success alone does not establish precise clock grounding.

\section{Construction and Quality Control}
\label{sec:construction}
\begin{figure}[t]
    \centering
    \includegraphics[width=\textwidth]{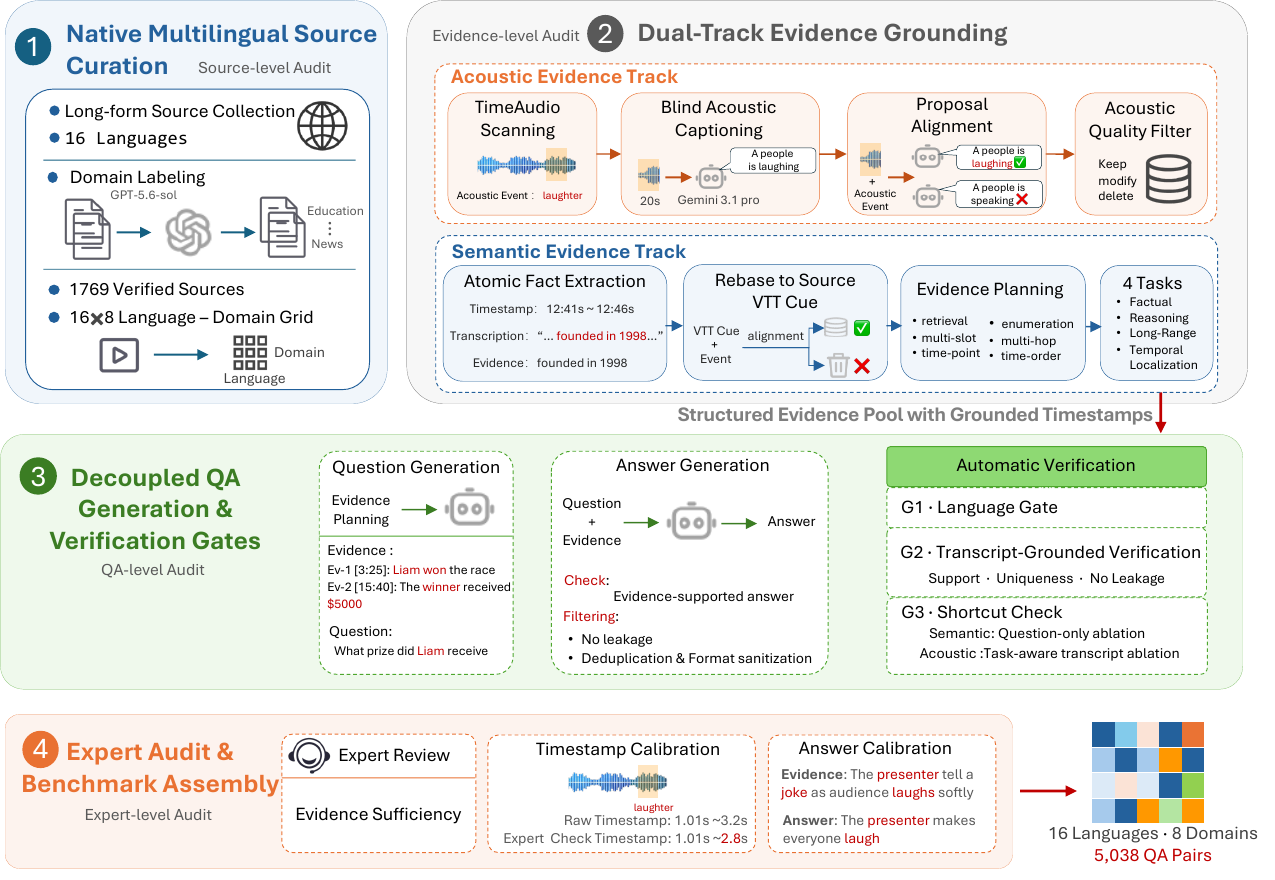}
    \caption{
    Construction pipeline of MuLA-Bench. Native multilingual sources are
    organized over the Language--Domain grid before semantic and acoustic
    evidence are grounded in parallel. Candidate QA pairs are generated from
    structured evidence, screened by automatic verification and shortcut tests,
    and finally audited against localized source evidence before release.
    }
    \label{fig:construction}
\end{figure}

MuLA-Bench uses an evidence-first pipeline: source curation, dual-track evidence
grounding, decoupled QA generation, and automatic and expert verification.
Figure~\ref{fig:construction} and Appendix~\ref{app:construction} give the
complete workflow and frozen settings.

\paragraph{Source and evidence grounding.}
Sources require usable audio and matching-language subtitles. Metadata flags
possible language mismatch or dubbing for inspection; confirmed invalid sources
and known overlap are removed. Semantic mining converts timestamped subtitle
cues into atomic facts and rebases each fact to its supporting source cue.
Weak alignments are discarded before evidence planning. For acoustic evidence,
TimeAudio \citep{wang2026timeaudio} proposes events across the full recording.
Up to 12 selected windows per source undergo two-pass Gemini 3.1 Pro
verification: proposal-hidden observation followed by proposal alignment.
Visual context may assist source verification, but retained events must be
audible. Metric temporal questions additionally require precise timing.

For acoustic events, the first verification pass records what is observable
without seeing the proposed label. The second pass aligns the proposal to this
observation using keep, modify, or delete decisions. Only sufficiently
supported events enter the question pool, and every cited event undergoes a
further localized audibility check. This concentrates expensive verification
on selected windows while preserving a full-recording proposal stage; it does
not claim exhaustive human or Gemini annotation of every sound.

\paragraph{QA generation and automatic gates.}
One call generates a question from a structured evidence plan; a separate
fresh-context call derives its answer. Items undergo language, grounding,
uniqueness, leakage, question-only, and timestamp checks. Acoustic events are
rechecked in localized audio, followed by a task-aware transcript-only test.
Factual, temporal, and long-range items are rejected if the full transcript
already answers them; reasoning items are rejected if speech local to the
acoustic event already states the target conclusion. This preserves the
intended contribution of sound without injecting target events.

The question writer receives no prewritten reference answer; a fresh-context
call derives the answer from the generated question and evidence. Language,
transcript-grounding, and question-only gates check validity. Repairable
candidates receive one correction and reverification; fatal grounding errors
are dropped. Packing enforces source-supported numerical values, multiple
evidence pieces for reasoning, and the semantic long-range rule.

\paragraph{Expert review.}
All 5,038 released questions receive final review by one paid expert for their
language, with 16 experts in total. Reviewers inspect the question, answer,
evidence, and localized audio, correct event timing where needed, and repair or
replace defective items. Exhaustive questions also receive a global transcript
check. More than 90\% of final candidates are accepted without replacement.
This is full-set quality control, distinct from the sampled judge-reliability
audit below. Review procedures and available rejection statistics are in
Appendix~\ref{app:construction}.

Experts can accept, repair, or replace a candidate after inspecting its cited
region. For temporal items, they correct the reference event and time against
audio playback; the automatic construction timestamp is not treated as final
human ground truth. The protocol uses one assigned expert per language, so the
full-set review is a quality-control process rather than an inter-annotator
agreement study. This distinction also separates dataset review from the
independent question of whether the automatic evaluator applies its rubric
consistently.

\section{Evaluation Protocol}
\label{sec:experiments}\label{sec:model_protocol}
\paragraph{Models and inputs.}
Table~\ref{tab:main_results} evaluates seven hosted and three open-weight
systems. Models receive the source-language question and canonical audio,
without transcripts, video, or gold evidence. The canonical input is the full
recording except for four predefined post-3h items using a fixed later-stage
three-hour segment. Local models receive the longest prefix permitted by their
native context limits, never a gold-centered crop. Missing, invalid, or
noncompliant responses score zero under the fixed 5,038-item denominator.
Appendix~\ref{app:evaluation_protocol} reports model settings and coverage.

In the frozen protocol, input limits are 3,276.8\,s for MOSS, 2,400\,s for
Qwen3-Omni, and 1,800\,s for Audio Flamingo Next. They therefore receive
different-length prefixes under the same input policy. The two additional
open-weight models return valid responses for 5,037 and 5,036 of 5,038 items,
respectively. High response coverage does not imply access to all annotated
evidence: their scores measure the complete system, including native context
limits, rather than answer quality conditional on observing the gold region.

\paragraph{Scoring and reliability.}
A binary rubric routes answers to single-fact, exhaustive-list, multi-slot,
reasoning, time-point, or time-order criteria. GPT-5.6 Sol checks semantic
agreement; deterministic guards enforce structural and temporal constraints.
Time-point scoring requires both the correct event and clock alignment within
$\pm3$ seconds, with explicit short-interval rules. Only final committed
answers are scored. Human re-scoring agrees with the evaluator on 114 of 117
sampled responses (97.4\%). Full rubrics, temporal matching, and audit scope
appear in Appendix~\ref{app:task_scoring}.

Exhaustive lists require all members, multi-slot answers require every mandatory
slot, and reasoning answers require the target conclusion and relation.
Time-order scoring checks sequence rather than a clock value. These
route-specific contracts prevent a lenient single-fact criterion from being
applied to stricter requests.

\paragraph{Diagnostic cohort.}
Table~\ref{tab:main_results} and Figure~\ref{fig:evidence_task} report all ten
models individually. Figures~\ref{fig:language_domain}
and~\ref{fig:distance_operation} retain the original eight-model cohort: the
seven hosted systems and MOSS. Holding this cohort fixed makes the pooled
residuals, position profiles, and regression estimates refer to the same model
population. These diagnostics are scoped to that cohort. Statistical
specifications are in Appendix~\ref{app:statistics}.

\FloatBarrier
\section{Results and Diagnostic Findings}
\label{sec:results}\label{sec:analysis}
\subsection{Overall Results}
\begin{table}[!htbp]
\centering
\caption{Accuracy (\%) on the fixed benchmark and slice denominators. Missing
responses score zero. Best and second-best results are bold and underlined.}
\label{tab:main_results}
\small
\setlength{\tabcolsep}{4.0pt}
\renewcommand{\arraystretch}{1.07}
\begin{tabular*}{\textwidth}{@{\extracolsep{\fill}}lrrrrrrr@{}}
\toprule
& & \multicolumn{2}{c}{\textbf{Evidence}} & \multicolumn{4}{c}{\textbf{Task}} \\
\cmidrule(lr){3-4}\cmidrule(lr){5-8}
\textbf{Model} & \textbf{Overall} & \textbf{Sem.} & \textbf{Aco.} &
\textbf{Factual} & \textbf{Reason.} & \textbf{Temporal} & \textbf{Long-range} \\
\midrule
\multicolumn{8}{l}{\textit{Hosted / API models}} \\
Gemini 3.8 Flash        & \textbf{73.40} & \textbf{79.06} & \textbf{55.26} & \textbf{71.29} & \textbf{72.51} & \textbf{66.67} & \textbf{85.30} \\
Gemini 3.7 Flash        & \underline{73.12} & \underline{78.98} & \underline{54.34} & \underline{70.81} & \textbf{72.51} & \underline{66.37} & \underline{85.10} \\
Gemini 3.1 Pro          & 61.31 & 72.14 & 26.63 & 56.43 & 68.85 & 39.58 & 83.99 \\
Qwen3.5-Omni-Plus       & 60.92 & 66.80 & 42.07 & 61.77 & 60.73 & 40.58 & 80.16 \\
Qwen3.5-Omni-Flash      & 46.55 & 52.45 & 27.63 & 47.96 & 44.42 & 27.68 & 65.46 \\
Doubao seed 2.0 Lite         & 56.91 & 62.86 & 37.81 & 59.97 & 60.30 & 26.79 & 77.74 \\
Muse Spark 1.2          & 33.98 & 36.88 & 24.71 & 37.39 & 33.86 & 13.19 & 48.74 \\
\midrule
\multicolumn{8}{l}{\textit{Open-weight models}} \\
MOSS-Audio-8B-Thinking  & 13.48 & 13.59 & 13.11 & 14.38 & 14.49 & 6.75 & 17.42 \\
Qwen3-Omni-30B-A3B-Instruct & 24.91 & 29.06 & 11.60 & 28.29 & 23.56 & 11.11 & 34.04 \\
Audio Flamingo Next 8B & 13.60 & 15.29 & 8.18 & 16.66 & 12.30 & 5.26 & 17.72 \\
\bottomrule
\end{tabular*}
\end{table}
Gemini 3.8 Flash and 3.7 Flash achieve 73.40\% and 73.12\% overall accuracy.
All ten models perform best on long-range retrieval and worse on acoustic
than semantic questions in aggregate. Qwen3-Omni-30B reaches 24.91\%, compared
with 13.60\% for Audio Flamingo Next and 13.48\% for MOSS; these scores include
native context constraints. Similar totals need not imply similar skills:
Gemini 3.1 Pro and Qwen3.5-Omni-Plus differ by only 0.40 points overall, yet
Pro leads on time-order and Plus on time-point and acoustic factual questions.
Complete evidence--task results are in Appendix~\ref{app:additional_results}.

\subsection{Conditional Multilingual Difficulty}
\label{sec:conditional_language_rankings}
\begin{figure}[t]
\centering
\includegraphics[width=\textwidth]{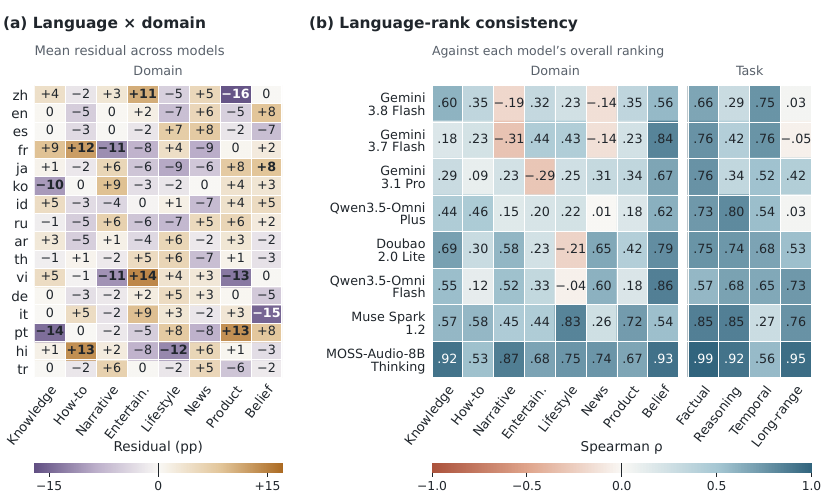}
\caption{Conditional language difficulty in the eight-model cohort.
(a) Mean semantic Language$\times$Domain residual (pp); bold values have
paired-bootstrap 95\% CIs excluding zero. Text color ensures contrast.
(b) Spearman correlation of conditional and overall language rankings.
Higher correlation means more consistent rankings, not higher accuracy.}
\label{fig:language_domain}
\end{figure}

On the balanced semantic grid, we remove each model's language and domain
marginals from its cell accuracy:
\begin{equation}
R_{m,\ell,d}=A_{m,\ell,d}-A_{m,\ell}-A_{m,d}+A_m.
\label{eq:ld_residual}
\end{equation}
Figure~\ref{fig:language_domain}(a) averages these residuals over the fixed
cohort. Vietnamese changes from $+14.2$ pp in Entertainment to $-13.3$ pp in
Product, while Chinese reaches $-16.4$ pp in Product. Such departures from the
additive expectation show that a language's relative difficulty depends on
the domain.

The effect extends to relative rankings (Figure~\ref{fig:language_domain}(b)).
Gemini 3.7 Flash's Narrative ranking correlates negatively with its overall
language ranking ($\rho=-0.31$); Gemini 3.1 Pro reaches $\rho=-0.29$ for
Entertainment. For Gemini 3.8 Flash, Japanese moves from first overall to tenth
on temporal questions, whereas Hindi moves from eighth overall to first on
long-range retrieval. These are benchmark-conditioned rankings, rather than a
single ordering intrinsic to the languages.

Here the residual and ranking panels address complementary questions. The
residual asks whether a specific language--domain combination departs from the
model's additive expectation; the ranking panel asks whether a conditional
slice preserves the relative ordering of languages. Averaging residuals does
not imply that all models share the same interaction pattern, as the
cross-model variation in Appendix~\ref{app:language_domain_interaction} shows.
Likewise, MOSS's comparatively stable rankings do not imply strong absolute
performance: its overall accuracy is 13.48\%. These two views distinguish
conditional difficulty from both model strength and a global language ranking.

\subsection{Operation-Dependent Evidence Gaps}
\begin{figure}[t]
\centering
\includegraphics[width=\textwidth]{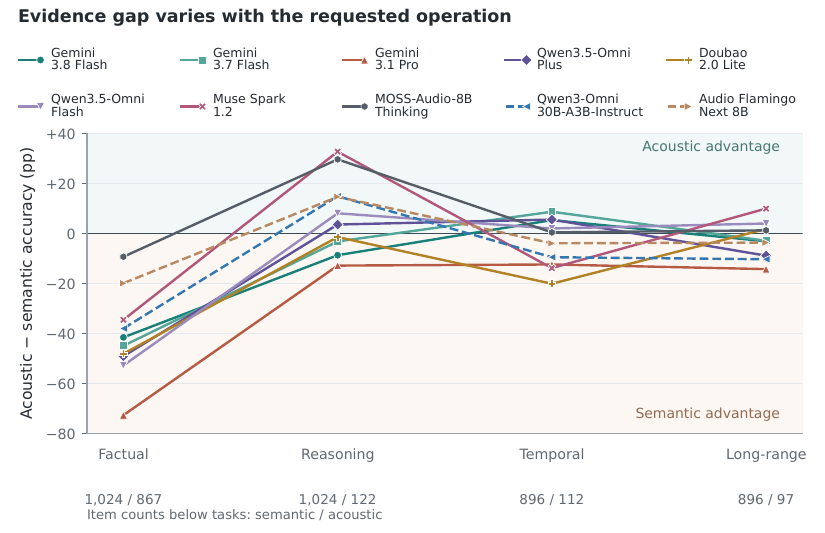}
\caption{Evidence gaps across operations for ten models.
$\Delta=100(\mathrm{Acc}_{\mathrm{aco}}-\mathrm{Acc}_{\mathrm{sem}})$;
negative values favor semantic evidence. All factual gaps are negative, while
other operations show mixed directions. Counts give semantic/acoustic items;
lines connect categorical tasks to show each model's profile.}
\label{fig:evidence_task}
\end{figure}

Figure~\ref{fig:evidence_task} shows acoustic-minus-semantic accuracy for all
ten models. Factual gaps are uniformly negative, from $-9.3$ pp for MOSS to
$-72.7$ pp for Gemini 3.1 Pro. In contrast, reasoning gaps reverse for six
models, including both additional open-weight systems, whose acoustic
advantages are about 15 pp. Temporal gaps are mixed, and long-range gaps are
generally smaller than factual gaps. Acoustic difficulty therefore depends on
the requested operation. These are descriptive comparisons of distinct item
subsets, not paired substitutions of evidence; the smaller acoustic samples
are shown in the figure.

The change of sign is central to this result. Gemini 3.1 Pro's factual accuracy
is 89.75\% on semantic items but 17.07\% on acoustic items. MOSS, despite much
lower overall performance, scores 11.33\% on semantic reasoning and 40.98\%
on acoustic reasoning. Thus the aggregate disadvantage of acoustic questions
coexists with operation-specific advantages. This pattern supports reporting
the evidence--operation profile alongside a total score. Because the two tracks
contain different questions and the acoustic task distribution is uneven, the
profile describes an interaction in observed performance, not the causal effect
of replacing spoken evidence with sound in an otherwise identical item.

\subsection{Evidence Distance}
\label{sec:distance_operation}
\begin{figure}[t]
\centering
\includegraphics[width=\textwidth]{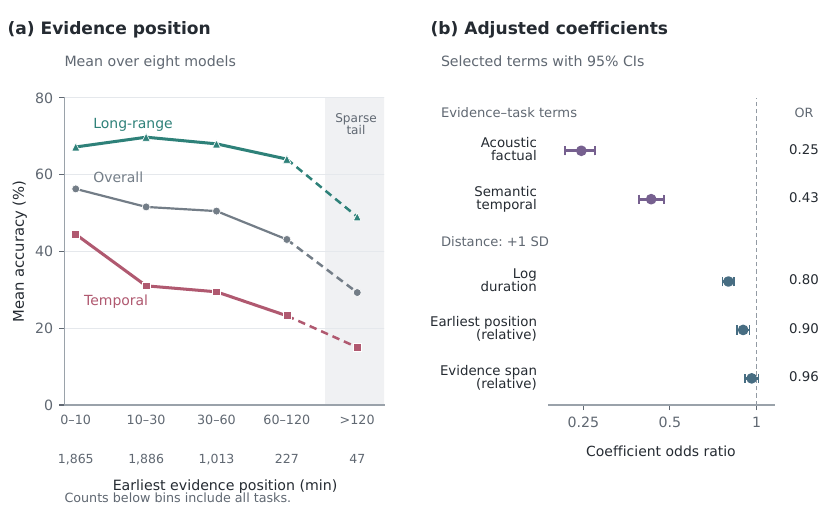}
\caption{Distance is not a uniform difficulty axis (eight-model cohort).
(a) Mean accuracy by earliest evidence position; counts include all tasks.
The dashed, shaded tail is sparse. (b) Adjusted ORs with 95\% CIs from 400
question-clustered bootstrap samples. Category contrasts use semantic factual
as reference; distance contrasts are +1 SD. The OR axis is logarithmic.}
\label{fig:distance_operation}
\end{figure}

Distance profiles differ by operation in the eight-model cohort
(Figure~\ref{fig:distance_operation}(a)). Across earliest-evidence bins from
0--10 through 60--120 minutes, long-range accuracy remains between 64.0\% and
69.7\%, while temporal accuracy falls from 44.4\% to 23.2\%. Both decline in
the sparse $>120$-minute tail ($n=47$ across tasks), to 49.0\% and 15.0\%.
Near-stability thus applies to the first four bins, not the full duration range.

A logistic model controlling for model, language, domain, joint evidence--task
category, and standardized distance/count variables complements these trends.
Acoustic factual and semantic temporal questions have adjusted odds ratios
of 0.096 and 0.147 relative to semantic factual questions
(Figure~\ref{fig:distance_operation}(b)). A +1 SD increase in log duration,
normalized earliest position, or span gives ORs of 0.798, 0.897, and 0.961;
the span CI crosses one. These are descriptive conditional associations,
not causal effects or a universal ranking of predictor importance.

The fit uses 40,304 question--model outcomes and one joint evidence--task
category variable, with semantic factual as the reference. It includes evidence
count alongside the three distance measures; it does not redundantly add
separate evidence and task indicators. Each bootstrap sample resamples
questions while retaining their eight model outcomes. This specification makes
the displayed category contrasts identifiable and preserves within-question
dependence when estimating their intervals. The raw curves and adjusted
associations together show why distance should be interpreted with the
requested operation, while the sparse late tail limits conclusions about the
most distant evidence.

\subsection{Event Understanding versus Clock Grounding}
Temporal questions distinguish ordering events from aligning them to physical
time. Several systems perform much better on ordering than time-point
localization (Table~\ref{tab:temporal_diagnostic}). The diagnostic slices contain 386 time-order and 567 time-point questions. The Gemini Flash variants
have a smaller gap, so the distinction is model-dependent.

\begin{wraptable}[8]{r}{0.46\textwidth}
\vspace{-12pt}
\centering
\caption{\small Temporal task accuracy (\%) for three illustrative models.}
\label{tab:temporal_diagnostic}
\vspace{2pt}

\small
\setlength{\tabcolsep}{7pt}
\renewcommand{\arraystretch}{1.08}
\begin{tabular}{lrr}
\toprule
Model & Order & Point \\
\midrule
Gemini 3.1 Pro       & 59.6 & 22.9 \\
Doubao seed 2.0 Lite & 48.7 &  8.1 \\
Muse Spark 1.2       & 26.4 &  1.1 \\
\bottomrule
\end{tabular}
\vspace{-5pt}
\end{wraptable}

Response-level cases further separate event recovery from clock alignment.
For a Turkish item asking when a stitch count rises from 63 to 69, several
models recover that event but report 15:57, 16:26, or 14:32 instead of 16:06.
The semantic event is available in these answers, while its physical time is
incorrect. Such errors motivate temporal grounding diagnostics beyond either
overall accuracy or event order. Appendix~\ref{app:cases} gives additional
qualitative cases; they illustrate failure types rather than estimate their
frequencies.

\section{Conclusion}
MuLA-Bench combines native multilingual long-form recordings, controlled
semantic question allocation, and natural acoustic evidence with multi-tier
auditing. Its evaluation reveals conditional language rankings,
operation-dependent evidence gaps, and failures to align understood events to
the recording clock.

A practical implication is to separate retrieving content, using natural
acoustic evidence, and aligning events to time when evaluating long-context
audio systems. A longer admissible input alone does not establish reliability
on all three. MuLA-Bench provides controlled comparisons and localized natural
evidence for diagnosing these differences across languages.

\bibliographystyle{iclr2027_conference}
\bibliography{lode_all_refs}

@inproceedings{yang2024airbench,
  title     = {{AIR}-Bench: Benchmarking Large Audio-Language Models via Generative Comprehension},
  author    = {Yang, Qian and Xu, Jin and Liu, Wenrui and Chu, Yunfei and Jiang, Ziyue and Zhou, Xiaohuan and Leng, Yichong and Lv, Yuanjun and Zhao, Zhou and Zhou, Chang and Zhou, Jingren},
  booktitle = {Proceedings of the 62nd Annual Meeting of the Association for Computational Linguistics (Volume 1: Long Papers)},
  pages     = {1979--1998},
  year      = {2024},
  month     = aug,
  address   = {Bangkok, Thailand},
  publisher = {Association for Computational Linguistics},
  doi       = {10.18653/v1/2024.acl-long.109},
  url       = {https://aclanthology.org/2024.acl-long.109/}
}

@inproceedings{wang2025audiobench,
  title     = {{AudioBench}: A Universal Benchmark for Audio Large Language Models},
  author    = {Wang, Bin and Zou, Xunlong and Lin, Geyu and Sun, Shuo and Liu, Zhuohan and Zhang, Wenyu and Liu, Zhengyuan and Aw, AiTi and Chen, Nancy F.},
  booktitle = {Proceedings of the 2025 Conference of the Nations of the Americas Chapter of the Association for Computational Linguistics: Human Language Technologies (Volume 1: Long Papers)},
  pages     = {4297--4316},
  year      = {2025},
  month     = apr,
  address   = {Albuquerque, New Mexico},
  publisher = {Association for Computational Linguistics},
  doi       = {10.18653/v1/2025.naacl-long.218},
  url       = {https://aclanthology.org/2025.naacl-long.218/}
}

@inproceedings{sakshi2025mmau,
  title     = {{MMAU}: A Massive Multi-Task Audio Understanding and Reasoning Benchmark},
  author    = {Sakshi, S and Tyagi, Utkarsh and Kumar, Sonal and Seth, Ashish and Selvakumar, Ramaneswaran and Nieto, Oriol and Duraiswami, Ramani and Ghosh, Sreyan and Manocha, Dinesh},
  booktitle = {International Conference on Learning Representations},
  year      = {2025},
  url       = {https://proceedings.iclr.cc/paper_files/paper/2025/hash/d36f208919582785db965fe648b9fe59-Abstract-Conference.html}
}

@inproceedings{ma2025mmar,
  title     = {{MMAR}: A Challenging Benchmark for Deep Reasoning in Speech, Audio, Music, and Their Mix},
  author    = {Ma, Ziyang and Ma, Yinghao and Zhu, Yanqiao and Yang, Chen and Chao, Yi-Wen and Xu, Ruiyang and Chen, Wenxi and Chen, Yuanzhe and Chen, Zhuo and Cong, Jian and Li, Kai and Li, Keliang and Li, Siyou and Li, Xinfeng and Li, Xiquan and Lian, Zheng and Liang, Yuzhe and Liu, Minghao and Niu, Zhikang and Wang, Tianrui and Wang, Yuping and Wang, Yuxuan and Wu, Yihao and Yang, Guanrou and Yu, Jianwei and Yuan, Ruibin and Zheng, Zhisheng and Zhou, Ziya and Zhu, Haina and Xue, Wei and Benetos, Emmanouil and Yu, Kai and Chng, Eng-Siong and Chen, Xie},
  booktitle = {Advances in Neural Information Processing Systems},
  volume    = {38},
  year      = {2025},
  url       = {https://proceedings.neurips.cc/paper_files/paper/2025/hash/610a7d6507d55be70c6d13d0b663227d-Abstract-Datasets_and_Benchmarks_Track.html}
}

@misc{ahia2025blab,
  title         = {{BLAB}: Brutally Long Audio Bench},
  author        = {Ahia, Orevaoghene and Bartelds, Martijn and Ahuja, Kabir and Gonen, Hila and Hofmann, Valentin and Arora, Siddhant and Li, Shuyue Stella and Puttagunta, Vishal and Adeyemi, Mofetoluwa and Buchireddy, Charishma and Walls, Ben and Bennett, Noah and Watanabe, Shinji and Smith, Noah A. and Tsvetkov, Yulia and Kumar, Sachin},
  year          = {2025},
  eprint        = {2505.03054},
  archivePrefix = {arXiv},
  primaryClass  = {cs.AI},
  doi           = {10.48550/arXiv.2505.03054},
  url           = {https://arxiv.org/abs/2505.03054}
}

@misc{yang2026longspeech,
  title         = {{LongSpeech}: A Scalable Benchmark for Transcription, Translation and Understanding in Long Speech},
  author        = {Yang, Fei and Ni, Xuanfan and Yang, Renyi and Geng, Jiahui and Li, Qing and Lyu, Chenyang and Du, Yichao and Wang, Longyue and Luo, Weihua and Zhang, Kaifu},
  year          = {2026},
  eprint        = {2601.13539},
  archivePrefix = {arXiv},
  primaryClass  = {cs.SD},
  doi           = {10.48550/arXiv.2601.13539},
  url           = {https://arxiv.org/abs/2601.13539}
}

@misc{luo2026chronosaudio,
  title         = {{ChronosAudio}: A Comprehensive Long-Audio Benchmark for Evaluating Audio-Large Language Models},
  author        = {Luo, Kaiwen and Lin, Liang and Zhang, Yibo and Aloqaily, Moayad and Tao, Jialiang and Wang, Dexian and Zhou, Zhenhong and Zhang, Junwei and Wang, Kun and Sun, Li and Wen, Qingsong},
  year          = {2026},
  eprint        = {2601.04876},
  archivePrefix = {arXiv},
  primaryClass  = {cs.SD},
  doi           = {10.48550/arXiv.2601.04876},
  url           = {https://arxiv.org/abs/2601.04876}
}

@misc{huang2026audiospan,
  title         = {{AudioSpan}: Spanning the Duration and Depth of Audio Comprehension},
  author        = {Huang, Wen and Chu, Yunfei and Gao, Meng and He, Haolin and Xu, Jin},
  year          = {2026},
  eprint        = {2608.26431},
  archivePrefix = {arXiv},
  primaryClass  = {cs.SD},
  doi           = {10.48550/arXiv.2608.26431},
  url           = {https://arxiv.org/abs/2608.26431}
}

@inproceedings{gao2025adubench,
  title     = {Benchmarking Open-ended Audio Dialogue Understanding for Large Audio-Language Models},
  author    = {Gao, Kuofeng and Xia, Shu-Tao and Xu, Ke and Torr, Philip and Gu, Jindong},
  booktitle = {Proceedings of the 63rd Annual Meeting of the Association for Computational Linguistics (Volume 1: Long Papers)},
  pages     = {4763--4784},
  year      = {2025},
  month     = jul,
  address   = {Vienna, Austria},
  publisher = {Association for Computational Linguistics},
  doi       = {10.18653/v1/2025.acl-long.237},
  url       = {https://aclanthology.org/2025.acl-long.237/}
}

@inproceedings{iyer2026scenebench,
  title     = {{SCENEBench}: An Audio Understanding Benchmark Grounded in Assistive and Industrial Use Cases},
  author    = {Iyer, Laya and Wang, Angelina and Koyejo, Sanmi},
  booktitle = {Proceedings of the 19th Conference of the European Chapter of the Association for Computational Linguistics (Volume 1: Long Papers)},
  pages     = {7123--7137},
  year      = {2026},
  month     = mar,
  address   = {Rabat, Morocco},
  publisher = {Association for Computational Linguistics},
  doi       = {10.18653/v1/2026.eacl-long.335},
  url       = {https://aclanthology.org/2026.eacl-long.335/}
}

@inproceedings{wang2026timeaudio,
  title     = {Listening Between the Frames: Bridging Temporal Gaps in Large Audio-Language Models},
  author    = {Wang, Hualei and Li, Yiming and Ma, Shuo and Liu, Hong and Wang, Xiangdong},
  booktitle = {Proceedings of the AAAI Conference on Artificial Intelligence},
  volume    = {40},
  number    = {31},
  pages     = {26233--26241},
  year      = {2026},
  doi       = {10.1609/aaai.v40i31.39827},
  url       = {https://ojs.aaai.org/index.php/AAAI/article/view/39827}
}

@article{ye2026voicegiraffe,
  title   = {VoiceGiraffe: A Benchmark for Extreme Long-Context Audio-Language Understanding},
  author  = {Ye, Jashin and Wang, Dongxiao and Ye, Yixuan and Zhou, Sashuai and Lin, Weihuang and Han, Mingyang and Wang, Kunpeng and Yuan, Zeyu and Li, Boyu and Shi, Haoxiang and Shu, Jingchen and Song, Jun and Zheng, Bo},
  journal = {arXiv preprint arXiv:2605.27976},
  year    = {2026}
}

\appendix
\setcounter{figure}{0}
\renewcommand{\thefigure}{A\arabic{figure}}
\renewcommand{\theHfigure}{appendix.\arabic{figure}}
\FloatBarrier
\section{Dataset Specification and Composition}
\label{app:specification}
MuLA-Bench includes 16 languages, eight domains, and two evidence tracks.
The semantic allocation is balanced at the question level; source counts and
natural acoustic event availability need not be balanced.
\subsection{Language Coverage and Metadata}
\label{app:language_metadata}

Table~\ref{tab:language_metadata} summarizes the coverage dimensions used when
selecting the 16 target languages. Region labels indicate broad areas of
established use rather than assigning a language to a single culture. Speaker
scale follows Ethnologue's \emph{All Users} notion, which combines first- and
second-language users; we report coarse ranges rather than licensed exact
counts.\footnote{\url{https://www.ethnologue.com/}} These metadata are
descriptive and are not used to infer model performance from language-resource
level.

\begin{table*}[!htbp]
\centering
\caption{Geographic, linguistic, script, and speaker-scale coverage of the 16
target languages. Speaker scale is based on Ethnologue's global L1+L2
(\emph{All Users}) population concept and is reported only in coarse ranges.}
\label{tab:language_metadata}
\small
\setlength{\tabcolsep}{4.5pt}
\renewcommand{\arraystretch}{1.06}
\begin{tabularx}{\textwidth}{@{}p{.12\textwidth}p{.045\textwidth}p{.19\textwidth}p{.225\textwidth}X@{}}
\toprule
\textbf{Language} & \textbf{ISO} & \textbf{Broad region} &
\textbf{Language family} & \textbf{Primary script(s) / speaker scale} \\
\midrule
English     & en & Europe / global                & Indo-European (Germanic)   & Latin / $>1$B \\
Chinese     & zh & East Asia                      & Sino-Tibetan (Sinitic)     & Han / $>1$B \\
Hindi       & hi & South Asia                     & Indo-European (Indo-Aryan) & Devanagari / 100M--1B \\
Spanish     & es & Europe / Americas              & Indo-European (Romance)    & Latin / 100M--1B \\
French      & fr & Europe / Africa / global       & Indo-European (Romance)    & Latin / 100M--1B \\
Arabic      & ar & Middle East / North Africa     & Afro-Asiatic (Semitic)     & Arabic / 100M--1B \\
Portuguese  & pt & Europe / Americas / Africa     & Indo-European (Romance)    & Latin / 100M--1B \\
Russian     & ru & Eastern Europe / N. Eurasia    & Indo-European (Slavic)     & Cyrillic / 100M--1B \\
Indonesian  & id & Southeast Asia                 & Austronesian               & Latin / 100M--1B \\
German      & de & Europe                         & Indo-European (Germanic)   & Latin / 100M--1B \\
Japanese    & ja & East Asia                      & Japonic                    & Kanji + kana / 100M--1B \\
Vietnamese  & vi & Southeast Asia                 & Austroasiatic              & Latin / 50M--100M \\
Turkish     & tr & West Asia / Southeast Europe   & Turkic                     & Latin / 50M--100M \\
Korean      & ko & East Asia                      & Koreanic                   & Hangul / 50M--100M \\
Italian     & it & Europe                         & Indo-European (Romance)    & Latin / 50M--100M \\
Thai        & th & Southeast Asia                 & Kra-Dai                    & Thai / 50M--100M \\
\bottomrule
\end{tabularx}
\end{table*}

\subsection{Domain Taxonomy and Allocation}
The eight domains are Knowledge \& Education, Skill \& Tutorial,
Narrative \& Storytelling, Entertainment \& Performance,
Lifestyle \& Physical Activities, News \& Commentary,
Product \& Review, and Belief \& Spiritual Culture.
The controlled semantic allocation is
\begin{equation}
    n_{\mathrm{sem}}(\ell,d)=30,
    \qquad
    |\mathcal{Q}_{\mathrm{sem}}|
    =\sum_{(\ell,d)\in\mathcal{G}} n_{\mathrm{sem}}(\ell,d)
    =16\times8\times30=3,840.
    \label{eq:semantic_grid}
\end{equation}
Each cell contains eight factual, eight reasoning, seven temporal, and seven
long-range questions. The acoustic track contains 1,198 verified questions
without an equal-cell constraint.
\subsection{Task Definitions and Edge Cases}

\paragraph{Factual/completeness.}
This task family covers requests for explicitly recoverable information,
including entities, values, events, and requested sets of facts. The family is
heterogeneous by design: most items require a single grounded fact, while a
smaller subset requires an exhaustive list or multiple mandatory slots. We
therefore use \emph{factual/completeness} as a task-family label rather than as
a claim that all items test exhaustive recall.

\paragraph{Reasoning.}
Reasoning questions require inference over evidence in the recording, including
causal, comparative, procedural, and multi-evidence relations. Questions that
can be answered from generic world knowledge without the recording are rejected
during construction.

\paragraph{Temporal localization.}
Temporal questions are separated into metric and ordinal forms. Metric
questions require identifying the correct event and aligning it to the physical
recording timeline; ordinal questions require only the correct event order. The
former are routed to \texttt{time\_point\_program}, the latter to
\texttt{time\_order}.

\paragraph{Long-range retrieval.}
For semantic questions, long-range status follows Eq.~\ref{eq:long_range} and
therefore requires either late evidence or a sufficiently wide separation among
topically coherent evidence points. This definition is intentionally based on
evidence position rather than nominal recording duration. Acoustic long-range
questions use verified acoustic-event evidence and are analyzed separately from
the semantic packing rule.

\paragraph{Formal semantic long-range rule.}
For evidence timestamps $\mathcal E_q=\{t_1,\ldots,t_m\}$ in seconds,
\begin{equation}
\mathrm{LR}(q)=
\begin{cases}
\mathbf{1}[t_1\geq600], & m=1,\\[3pt]
\mathbf{1}\!\left[(\max_i t_i-\min_i t_i\geq600)\;\lor\;(\min_i t_i\geq600)\right],
& m>1.
\end{cases}
\label{eq:long_range}
\end{equation}
This rule includes a single late fact; long-range does not imply that every
item requires multi-hop reasoning. Multiple evidence points must be topically
coherent. Acoustic long-range items are generated separately from verified
acoustic evidence.

\subsection{Source Distributions}
The collection contains 1,769 source recordings totaling 1,377.9 hours, with
median duration 36.8 minutes and mean duration 46.7 minutes. The proportion
lasting at least 30 minutes is 92.8\%. Figure~\ref{fig:dataset_composition}
shows the duration and source-coverage distributions.
\subsection{Task by Evidence Type}

\begin{table}[!htbp]
\centering
\caption{Question composition by task and evidence source.}
\label{tab:task_evidence_counts}
\small
\setlength{\tabcolsep}{6pt}
\renewcommand{\arraystretch}{1.08}
\begin{tabular}{lrrr}
\toprule
\textbf{Task} & \textbf{Semantic} & \textbf{Acoustic} & \textbf{Total} \\
\midrule
Factual / Completeness & 1,024 & 867 & 1,891 \\
Reasoning & 1,024 & 122 & 1,146 \\
Temporal Localization & 896 & 112 & 1,008 \\
Long-range Retrieval & 896 & 97 & 993 \\
\midrule
\textbf{Total} & \textbf{3,840} & \textbf{1,198} & \textbf{5,038} \\
\bottomrule
\end{tabular}
\end{table}

\subsection{Language--Domain Coverage}

The semantic track follows the same 30-question allocation in every one of the
128 Language--Domain cells. The acoustic track is intentionally less uniform:
39 cells contain fewer than 10 acoustic questions, 14 contain more than 10, and
one cell is empty because no usable naturally occurring acoustic evidence was
available under the release criteria. We preserve these differences rather than
fill them with cross-language substitutions or synthetic events.
\FloatBarrier
\section{Construction and Quality Assurance}
\label{app:construction}

This appendix expands the construction stages summarized in
Figure~\ref{fig:construction}. The goal is to make the released benchmark
traceable from source selection to final expert audit for the frozen 5,038-item pipeline.

\subsection{Pipeline Summary}

\begin{table*}[!htbp]
\centering
\caption{Operational summary of the frozen MuLA-Bench construction pipeline.}
\label{tab:construction_pipeline_summary}
\small
\setlength{\tabcolsep}{5pt}
\renewcommand{\arraystretch}{1.08}
\begin{tabular}{p{0.15\textwidth}p{0.17\textwidth}p{0.24\textwidth}p{0.34\textwidth}}
\toprule
\textbf{Stage} & \textbf{Model / tool} & \textbf{Primary input} & \textbf{Main acceptance criterion} \\
\midrule
Source curation & rules + GPT-5.6 Sol & source metadata, matching VTT & usable audio/subtitles, valid language, no known leakage or confirmed dub/overlay \\
Semantic evidence & GPT-5.6 Sol + VTT rebase & timestamped subtitle cues & atomic fact supported by a real source cue; weak rebase rejected \\
Acoustic evidence & TimeAudio + Gemini 3.1 Pro & full recording + selected local clips & event is audibly present after blind-listen/alignment verification \\
QA generation & GPT-5.6 Sol & structured evidence plan & question has one evidence-supported target; fresh-context answer agrees \\
Automatic audit & programmatic gates + GPT-5.6 Sol & candidate QA + transcript/evidence & language, grounding, uniqueness, no leakage, no unintended shortcut \\
Expert audit & language experts & localized audio, transcript, QA, timestamps & answer/evidence correct; temporal reference corrected when necessary \\
\bottomrule
\end{tabular}
\end{table*}

The shared GPT-5.6 Sol calls used by semantic evidence mining, evidence planning,
question writing, fresh-context answer derivation, and textual verification use
temperature $0$, high reasoning effort, and a maximum completion budget of
8,192 tokens in the frozen construction configuration.

\subsection{Source Curation and Domain Structuring}

A source enters construction with a catalog language code and a matching
language VTT file. Language is a recording-level label inherited by its
questions. Metadata terms suggesting dubbing or simultaneous interpretation are
used only to trigger inspection; hard exclusion requires direct evidence in the
subtitle/audio content, such as bilingual overlap or a mismatch between the
declared language and dominant subtitle content. Occasional code-switching is
not itself an exclusion criterion.

Domain assignment is performed over the transcript with GPT-5.6 Sol using an
eight-way closed taxonomy: knowledge, how-to, narrative, entertainment,
lifestyle, news, product, and belief/spiritual culture. The classifier chooses
a dominant source-level function rather than segment-level topics. Known
predecessor sources and leakage lists are excluded before QA generation. After
supplementation or relabeling, 19 released recordings contribute questions to
more than one question-level domain cell; this does not change the source-level
dominant-domain statistics shown in Figure~\ref{fig:dataset_composition}.

\subsection{Semantic Evidence Mining and Planning}

Semantic evidence is mined from VTT text in long slices and represented as
atomic, self-contained facts with timestamps and information types. The frozen
configuration processes up to 40 minutes per mining slice and targets roughly
8--15 high- or medium-information facts per 10 minutes, while allowing denser
segments to contribute more. Low-information filler is not counted toward the
usable-fact pool.

Each mined fact is rebased to the real VTT cue that supports it. A standard
semantic fact requires a rebase score of at least $2.5$; temporal plans use the
stricter threshold $3.5$. Facts that cannot be tied back to a sufficiently
confident source cue are discarded. This step is the source of semantic timing
in the released benchmark; MuLA-Bench does not use forced alignment to create
the semantic timestamps in the official 5,038-item release.

The evidence planner then creates structured plans over the retained facts.
Supported structures include retrieval, enumeration, multi-slot, multi-hop,
time-point, and time-order. Multi-hop plans must use at least two facts linked
by a concrete causal, comparative, temporal, procedural, or arithmetic relation;
plans spanning distant but unrelated statements are rejected. Long-range
semantic plans follow Eq.~\ref{eq:long_range}. Although the generation code can
also propose unanswerable and false-premise negatives, these plan kinds are
excluded by the frozen 5,038-item packer.

\subsection{Natural Acoustic Evidence Construction}

\paragraph{Full-recording proposal stage.}
TimeAudio scans the full audio using 20-s windows with a 20-s hop. The frozen
configuration uses the TimeAudio checkpoint built on Whisper-large-v2, BEATs,
and Vicuna-7B components and requests event descriptions with start and end
timestamps. This stage is exhaustive over the recording windows; the later
Gemini stage is not.

\paragraph{High-value window selection.}
Because multimodal re-listening is substantially more expensive, a heuristic
selector retains at most 12 candidate windows per recording. Event classes are
assigned fixed informativeness weights, with short distinctive effects,
applause, laughter, cheering, music, and other salient non-speech events ranked
above weak breath or background-noise categories. At the source level, the
selection score is
\begin{equation}
    S = W + 4N_{\mathrm{div}} + 2N_{\mathrm{high}},
    \label{eq:acoustic_window_score}
\end{equation}
where $W$ is the summed event weight, $N_{\mathrm{div}}$ counts distinct
nontrivial event classes, and $N_{\mathrm{high}}$ counts high-weight events.
The selector first promotes class diversity, then fills remaining slots with
high-value windows while avoiding redundant windows where possible.

\paragraph{Two-pass multimodal verification.}
Gemini 3.1 Pro verifies selected source clips in two passes. Pass~1 is blind to
the TimeAudio proposal: the model watches and listens to the clip and lists
non-speech events it can actually perceive. Visual context is used only to
support or rule out a possible source; the event must remain audible. Pass~2
then reveals the TimeAudio proposal and aligns each proposal to the blind
observation with \textit{keep}, \textit{modify}, or \textit{delete}. The
alignment prompt is deliberately conservative for laughter, applause, crying,
and crowd reactions, while requiring stronger evidence before retaining
specific ambient-source labels such as water, wind, or background music.

Post-processing converts the two-pass decisions into event records. Only events
with sufficient evidence strength enter the QA pool. Coarse in-window events
remain usable for non-temporal questions, whereas temporal questions require
precise timing metadata. Expert-rejected event IDs and confirmed invalid
acoustic sources are removed before harvesting.

\subsection{Decoupled Question and Answer Generation}

The question-writing stage receives a structured evidence plan and is explicitly
instructed to output the question only. It may see the intended evidence plan
but not a prewritten reference answer. Its prompt prohibits answer keywords,
source timestamps, explicit reasoning operations, and location hints, and
requires a single defensible answer grounded in the recording rather than world
knowledge.

A separate fresh-context call then receives the generated question together
with the supporting evidence and derives the answer. For semantic questions,
this answer is derived from transcript evidence; for acoustic questions, the
released gold is tied to the verified event-grounded evidence record rather
than to a Gemini full-recording teacher answer. The answer call and question
writer are the same model family but separate calls; MuLA-Bench does not use a
Claude/Gemini reciprocal-generation scheme in the frozen 5,038-item pipeline.

\subsection{Automatic Verification Gates}

The main semantic verifier is organized into three layers. \textbf{G1} is a
programmatic language gate over the generated question and answer. \textbf{G2}
checks the candidate against transcript context for unsupported claims, false
premises, multiple valid answers, answer or method leakage, omissions, and QA
mismatch. When the transcript fits the verifier budget it is supplied in full;
for longer transcripts, deterministic beginning/middle/end coverage is used
instead of claiming complete transcript visibility. Repairable failures such as
answer leakage or a narrow QA mismatch may be minimally repaired once and
reverified; fatal grounding defects are dropped. \textbf{G3} is a
question-only ablation: a solver is asked to answer without audio, transcript,
or cited evidence, and a second check removes the item if the guess materially
matches the reference.

After LLM verification, programmatic packer rules enforce additional invariants.
Questions must retain nonempty evidence; semantic numeric answers must be
supported by the source cue rather than only by a generated fact summary;
reasoning items require at least two evidence pieces; long-range items must
satisfy the corresponding temporal geometry; and semantic time-point golds must
remain close to their stored listening point. Standard semantic evidence uses a
minimum rebase score of $2.5$, while temporal items use $3.5$. These checks are
construction-time consistency guards and are distinct from the $\pm3$~s
evaluation tolerance.

\subsection{Task-Aware Acoustic Shortcut Tests}

Acoustic items receive a stricter but task-dependent text ablation. Let
$T_{\mathrm{full}}$ denote the full transcript and $T_{\mathrm{local}}$ the
speech local to the cited acoustic event. We reject an acoustic item $q$ when
\begin{equation}
\mathrm{Reject}_{\mathrm{aco}}(q)=
\begin{cases}
1, & \tau(q)\in\{\mathrm{fact},\mathrm{long},\mathrm{time}\}
     \;\land\; T_{\mathrm{full}} \Rightarrow a_q,\\[3pt]
1, & \tau(q)=\mathrm{reasoning}
     \;\land\; T_{\mathrm{local}} \Rightarrow a_q,\\[3pt]
0, & \text{otherwise.}
\end{cases}
\label{eq:acoustic_ablation}
\end{equation}
For factual/completeness, long-range, and temporal questions, the full
transcript must therefore be insufficient. For reasoning, the intended question
is different: the acoustic event should contribute a local premise. Distant
transcript passages that could be combined after the fact into a similar guess
do not make the sound redundant if the speech at the cited scene does not
already state the conclusion.

The implementation is two-stage. A transcript-only solver first declares
whether it can answer under the task-specific rule. If it declares the item
unanswerable, the candidate proceeds. If it declares the item answerable, a
second verifier compares the transcript-only answer to the reference; only a
confirmed material match is recorded as a shortcut and removed.

In addition, every cited acoustic event is re-presented in its localized source
window for an independent audibility check. A rejected or uncertain cited event
invalidates the corresponding evidence plan before expert review.

\subsection{Human Annotation Protocol}
\label{app:human_annotation}

\paragraph{Annotators and assignment.}
We use one paid language expert for each of the 16 target languages. Annotators
have relevant annotation experience and language credentials; we do not assume
native-speaker status. Each released item receives one final review by the
expert assigned to its language. This final pass is a quality-control audit,
not a multi-annotator agreement study.

\paragraph{Review interface and evidence access.}
For each candidate, reviewers have access to the question, reference answer,
transcript evidence, evidence timestamp(s), source link, and the corresponding
localized audio segment. Reviewers are required to listen to the localized
evidence region rather than judge from text alone. For acoustic questions, the
cited non-speech or paralinguistic event must be audibly present in that region.
For completeness-style questions that require exhaustive coverage, reviewers
additionally search the full VTT to check for relevant evidence elsewhere in
the recording.

\paragraph{Decision categories and resolution.}
The annotation interface records whether the item is correct or exhibits a
question problem, incorrect answer, incorrect sound, incorrect time,
unintelligible audio, or another issue. Reviewers check evidence sufficiency,
answer correctness, and, where applicable, acoustic-event and timestamp
correctness. Correct items are accepted. Repairable errors in the question,
answer, evidence, or timestamp are corrected; irreparable candidates are
replaced. All 5,038 released questions pass this final expert review, and more
than 90\% of the submitted final candidates are accepted without replacement.

\paragraph{Temporal calibration.}
Temporal review explicitly separates construction-time consistency, human
annotation, and evaluation tolerance. Semantic construction first applies a
coarse consistency gate between the proposed clock and its localized evidence
point. The expert then replays the localized region, identifies when the
referenced sound or spoken event actually occurs, and corrects the reference
time when necessary. Evaluation finally applies the deterministic temporal
matcher in Eq.~\ref{eq:time_score}; its $\pm3$~s tolerance is therefore a
scoring rule, not a substitute for timestamp annotation.

\subsection{Frozen Construction Configuration}

\begin{table*}[!htbp]
\centering
\caption{Key frozen configuration choices used to construct MuLA-Bench.}
\label{tab:construction_config}
\small
\setlength{\tabcolsep}{5pt}
\renewcommand{\arraystretch}{1.08}
\begin{tabularx}{\textwidth}{@{}p{.20\textwidth}p{.38\textwidth}X@{}}
\toprule
\textbf{Component} & \textbf{Frozen setting} & \textbf{Role} \\
\midrule
Shared LLM & GPT-5.6 Sol, temperature 0, high reasoning effort & evidence mining, planning, QA, textual verification \\
Semantic slice & 40 min maximum per mining slice & atomic-fact extraction \\
Semantic rebase & score $\geq2.5$; temporal $\geq3.5$ & source-cue grounding \\
TimeAudio sweep & 20-s window / 20-s hop & full-recording event proposals \\
Gemini verification & Gemini 3.1 Pro, at most 12 selected windows/source & blind listen + proposal alignment \\
Acoustic harvest & at most 10 per L$\times$D cell; at most 3 per source & preserve diversity without forcing sparse cells \\
Temporal evaluation & $\pm3$ s & final scoring tolerance only \\
\bottomrule
\end{tabularx}
\end{table*}

Appendix~\ref{app:prompts} specifies the information supplied to each model
call and the output decisions used by the pipeline. Programmatic gates remain
separate from the semantic decisions made by the models.

\subsection{Audit and Rejection Statistics}

Event-level expert auditing removes 56 rejected acoustic event IDs before final
question harvesting, and eight confirmed invalid acoustic sources are excluded.
The released benchmark contains 1,769 unique source recordings and 5,038
questions after all automatic gates and expert verification.
\FloatBarrier
\section{Evaluation and Statistical Protocol}
\label{app:evaluation_protocol}

\subsection{Frozen Model Pool, Snapshots, and Decoding}

Table~\ref{tab:eval_model_config} records the frozen model pool, exact model identifiers, and
inference settings. For hosted APIs, we use provider defaults unless otherwise
noted and do not explicitly set temperature, thinking mode, or a reasoning
budget. Local open-weight models are decoded greedily with the limits shown in
the table.

\begin{table*}[!htbp]
\centering
\caption{Frozen evaluation model snapshots and decoding settings.}
\label{tab:eval_model_config}
\small
\setlength{\tabcolsep}{4.8pt}
\renewcommand{\arraystretch}{1.07}
\begin{tabularx}{\textwidth}{@{}p{.205\textwidth}p{.32\textwidth}p{.22\textwidth}X@{}}
\toprule
\textbf{Paper name} & \textbf{Model identifier} & \textbf{Sampling / thinking} & \textbf{Output limit} \\
\midrule
Gemini 3.8 Flash & \nolinkurl{gemini-3.8-flash} & provider default & provider default \\
Gemini 3.7 Flash & \nolinkurl{gemini-3.7-flash} & provider default & provider default \\
Gemini 3.1 Pro & \nolinkurl{gemini-3.1-pro-preview} & provider default & provider default \\
Qwen3.5-Omni-Plus & \nolinkurl{qwen3.5-omni-plus} & provider default & provider default \\
Qwen3.5-Omni-Flash & \nolinkurl{qwen3.5-omni-flash} & provider default & provider default \\
Doubao seed 2.0 Lite & \makecell[l]{\texttt{doubao-seed-2-0-}\\\texttt{lite-260428}} & provider default & provider default \\
Muse Spark 1.2 & \texttt{muse-spark-1.2} & provider default & 8,192 tokens \\
MOSS-Audio-8B-Thinking & local checkpoint & greedy & 1,024 tokens \\
Qwen3-Omni-30B-A3B-Instruct & local checkpoint & greedy & thinker: 512 tokens \\
Audio Flamingo Next 8B & local checkpoint & greedy; repetition penalty 1.2 & 512 tokens \\
\bottomrule
\end{tabularx}
\end{table*}

\subsection{Question Prompt}

All systems receive the original benchmark question preceded by a fixed
language-specific wrapper equivalent to ``Please answer the following question
based on the full audio content.'' No additional system instruction,
chain-of-thought request, or JSON schema is added. This wrapper is translated
into each of the 16 target languages and is identical across models for a given
item. The source transcript, construction evidence, and video frames are never
provided at evaluation time.

\subsection{Canonical Input and Context Limits}

The canonical input is the full source audio, except for four predefined items
whose evidence occurs after the three-hour mark. These four use a fixed
later-stage three-hour segment containing all annotated evidence. Content-
preserving re-encoding or bitrate reduction is permitted; evidence-aware
cropping is not. Hosted models that cannot process the canonical input after
ordinary adaptation receive zero on the item after transient failures are
retried.

For local models with fixed native audio-context limits, we feed the longest
prefix permitted by the model, starting from the beginning of the canonical
input. The frozen limits are 3,276.8\,s for MOSS-Audio-8B-Thinking, 2,400\,s for
Qwen3-Omni-30B-A3B-Instruct, and 1,800\,s for Audio Flamingo Next 8B. For the
four predefined post-3h items, the prefix starts at the beginning of the
predefined later-stage segment, never at the gold evidence. Thus context-window
limitations remain part of the measured system behavior rather than being
hidden by oracle cropping.

\subsection{Fixed-Denominator Scoring and Coverage}

All headline and slice accuracies use the official number of benchmark items as
the denominator. For the full benchmark this is 5,038, regardless of whether a
model returns a valid response. After retrying transient network or server
errors, persistent input-limit failures, empty responses, and policy refusals
are scored as incorrect. We separately report
\emph{coverage}---the fraction of items that yield a non-empty valid response
under the prescribed input---as an operational diagnostic.

\begin{table}[!htbp]
\centering
\caption{Valid-response coverage for all ten models. Coverage does not
change the 5,038-item accuracy denominator.}
\label{tab:coverage}
\small
\setlength{\tabcolsep}{5pt}
\begin{tabular}{lrr}
\toprule
\textbf{Model} & \textbf{Valid / total} & \textbf{Coverage (\%)} \\
\midrule
Gemini 3.8 Flash & 5,036 / 5,038 & 99.96 \\
Gemini 3.7 Flash & 5,038 / 5,038 & 100.00 \\
Gemini 3.1 Pro & 5,038 / 5,038 & 100.00 \\
Qwen3.5-Omni-Plus & 5,033 / 5,038 & 99.90 \\
Doubao seed 2.0 Lite & 4,916 / 5,038 & 97.58 \\
Muse Spark 1.2 & 5,030 / 5,038 & 99.84 \\
MOSS-Audio-8B-Thinking & 5,038 / 5,038 & 100.00 \\
Qwen3.5-Omni-Flash & 4,759 / 5,038 & 94.46 \\
Qwen3-Omni-30B-A3B-Instruct & 5,037 / 5,038 & 99.98 \\
Audio Flamingo Next 8B & 5,036 / 5,038 & 99.96 \\
\bottomrule
\end{tabular}
\end{table}

\subsection{Evaluation Rubric Design}
\label{app:task_scoring}

Open-ended evaluation is organized around \emph{answer structure} rather than a
single holistic similarity score. Each item is mapped to one of six routes with
an explicit binary acceptance rubric. This separation matters because the same
benchmark task can require very different answer contracts: a factual item may
request one entity, an exhaustive set, or several mandatory slots, while a
temporal item may require either a physical clock value or only an ordering.
The routing therefore prevents a lenient single-fact criterion from being
reused for intrinsically stricter outputs.

We follow four design principles. First, \textbf{semantic tolerance}: harmless
paraphrases, translations, unit-equivalent expressions, and benign
transcription variants are accepted when identity remains unambiguous. Second,
\textbf{structural strictness}: exhaustive and multi-slot questions receive no
credit when a required element is missing. Third, \textbf{claim-level
reasoning}: reasoning responses are judged on the primary claims required by
the reference, not stylistic overlap or the presence of a particular chain of
thought. Fourth, \textbf{deterministic temporal grounding}: the LLM judge only
matches events to candidate times; final clock acceptance is decided by code.

\begin{table*}[!htbp]
\centering
\caption{Route-specific binary rubrics used for open-ended evaluation.}
\label{tab:judge_rules}
\small
\setlength{\tabcolsep}{5.5pt}
\renewcommand{\arraystretch}{1.10}
\begin{tabularx}{\textwidth}{@{}p{.235\textwidth}rX@{}}
\toprule
\textbf{Judge route} & \textbf{Count} & \textbf{Binary acceptance criterion} \\
\midrule
\texttt{single\_fact} & 2,458 & The final answer commits to the correct core grounded fact; benign paraphrase or transcription variants are allowed unless the question explicitly requires an exact string. \\
\texttt{exhaustive\_list} & 79 & Every required item in the reference set is present; any required omission is incorrect. \\
\texttt{multi\_slot} & 394 & Every mandatory slot is answered correctly and separately; one missing or incorrect slot makes the item incorrect. \\
\texttt{reasoning} & 1,145 & The final answer supports the primary reference conclusion and required claim(s); stylistic or intermediate-reasoning differences are ignored. \\
\makecell[l]{\texttt{time\_point}\\\texttt{\_program}} & 574 & The referenced event must match semantically and the predicted point/interval must pass the deterministic temporal matcher. \\
\texttt{time\_order} & 388 & The required event ordering must be correct; metric timestamp accuracy is not required. \\
\midrule
\textbf{Total} & \textbf{5,038} & \\
\bottomrule
\end{tabularx}
\end{table*}

\paragraph{Judge configuration.}
All semantic judge routes use GPT-5.6 Sol. The frozen evaluator uses low
reasoning effort for \texttt{single\_fact} and medium effort for
\texttt{exhaustive\_list}, \texttt{multi\_slot}, \texttt{reasoning},
\texttt{time\_point\_program}, and \texttt{time\_order}. The judge receives the
question, reference answer, and model prediction under the route-specific
rubric. Programmatic guards then enforce route-specific hard constraints.

\subsection{Deterministic Temporal Matching}

For metric temporal questions, the LLM judge identifies the event--time pairing
but does not decide the numeric tolerance. The programmatic matcher uses a
$3.0$\,s tolerance. Point--point predictions pass when the absolute difference
is at most $3.0$\,s. If the reference is an interval and the model predicts a
point, the point must fall inside the reference interval after expanding each
endpoint by $3.0$\,s. If the model predicts an interval, its width must be at
most $10.0$\,s; such a short interval passes only when it contains a reference
point (under the same tolerance) or overlaps a reference interval after
expansion. Wider predicted windows do not receive credit merely for covering
the answer. Gold spans of at least $0.5$\,s, or items explicitly typed as
intervals, are treated as reference intervals. \jtype{time-order} bypasses
these metric rules and scores sequence only.

\subsection{Final-Answer Policy}

Only the final committed answer is scored. Intermediate reasoning,
self-correction, or discarded candidates are ignored. If the final response
contains mutually competing unresolved answers, the item is marked incorrect.
A response that claims it cannot access the audio but then guesses from external
knowledge receives no credit unless the final committed answer independently
satisfies the route-specific criterion. Missing responses and persistent
inference failures are incorrect under the fixed-denominator protocol.

\subsection{Human--Judge Reliability Audit}

Because MuLA-Bench relies on an LLM judge for semantic equivalence rather than
string matching, we separately audit the evaluator against human review. We
manually re-score 117 sampled model responses using the same route-specific
criteria. The automatic evaluator agrees with human review on 114 of 117 cases
(97.4\%). We report this as \emph{agreement}, not as an estimate of absolute
judge accuracy, because the audit is a sampled reliability check rather than a
separately constructed adjudication benchmark. The audit is intended to verify
that the routed rubric behaves consistently with human application of the same
rules; it does not score or reward hidden chain-of-thought content.

\subsection{Scores and Statistical Analyses}
\label{app:statistics}
For a fixed evaluation subset $\mathcal S$, with binary item scores $s_q$,
\begin{equation}
    \mathrm{Acc}(\mathcal{S})
    =\frac{1}{|\mathcal{S}|}\sum_{q\in\mathcal{S}} s_q.
    \label{eq:accuracy}
\end{equation}
Missing and invalid outputs remain in the denominator. Wilson 95\% intervals
summarize individual model proportions. They describe uncertainty over the
benchmark questions and are not estimates over a random population of models.

\paragraph{Temporal score.}
Semantic event agreement and physical-clock matching are both required:
\begin{equation}
    s_q^{\mathrm{time}}
    =\mathbf{1}[\mathrm{event}(\hat a_q)=\mathrm{event}(a_q)]
     \cdot M_{\mathrm{time}}(\hat t_q,t_q^\star).
    \label{eq:time_score}
\end{equation}
Here $M_{\mathrm{time}}$ applies the point/interval rules above. Task families
and judge routes are distinct taxonomies; a task-conditioned slice need not
have the same denominator as an all-task judge-route count.

\paragraph{Language--Domain residuals and rankings.}
Each model's residual is computed from Eq.~\ref{eq:ld_residual} on the balanced
semantic track, then averaged across the fixed eight-model cohort. Bold cells
in Figure~\ref{fig:language_domain} use the paired-bootstrap 95\% intervals
from the frozen analysis. The overall language ranking uses all benchmark
questions; domain rankings use the semantic cell accuracies, and task rankings
use the corresponding task-specific accuracies. Spearman correlations measure
ranking agreement over the 16 languages. Cross-model residual variation is
shown in Figure~\ref{fig:language_domain_std}.

\paragraph{Distance profiles.}
Earliest evidence is measured in seconds from the recording origin.
Normalized position and span divide by source duration. Figure~\ref{fig:distance_operation}
groups items by earliest evidence position and averages model accuracies within
each slice. Counts displayed below bins include all tasks, not separate
temporal or long-range denominators. The $>120$-minute tail contains 47 items.

\paragraph{Identifiable logistic specification.}
The descriptive model uses 40,304 item--model observations (5,038 questions
across eight models). Its linear predictor is
\begin{equation}
\begin{split}
\operatorname{logit}\Pr(Y_{qm}=1)={}&\beta_0+\alpha_m+\lambda_{\ell(q)}+
\delta_{d(q)}+\gamma_{e(q),t(q)}\\
&+\boldsymbol\theta^\top\mathbf z_q,
\end{split}
\end{equation}
where $\mathbf z_q$ contains standardized log duration, normalized earliest
position, normalized evidence span, and evidence count. Joint evidence--task
categories use semantic factual as reference; model, language, and domain
references are Gemini 3.8 Flash, English, and Knowledge. No additional evidence
or task main-effect indicators are included. The fit is unpenalized, with
400 question-clustered bootstrap samples for 95\% intervals. Each sampled
question retains its eight model outcomes. Category contrasts and +1 SD
continuous shifts describe different comparisons; neither is interpreted as
causal variable importance.
\FloatBarrier
\section{Additional Results}
\label{app:additional_results}
The evidence--task and headline tables cover all ten models. Language/domain
breakdowns and pooled diagnostics use the fixed eight-model cohort.
In the compact tables, G3.8/G3.7/G3.1 denote Gemini 3.8 Flash/3.7 Flash/3.1 Pro;
Q-Plus/Q-Flash denote Qwen3.5-Omni-Plus/Flash; Spark denotes Muse Spark 1.2;
and MOSS denotes MOSS-Audio-8B-Thinking.

\subsection{Language and Domain Breakdowns}

\begin{table}[H]
\centering
\caption{Overall language accuracy (\%) over both evidence tracks, with official per-language counts.}
\label{tab:language_results}
\small
\setlength{\tabcolsep}{3pt}
\begin{tabular*}{\textwidth}{@{\extracolsep{\fill}}lrrrrrrrrr@{}}
\toprule
Lang. & $n$ & G3.8 & G3.7 & G3.1 & Q-Plus & Doubao & Q-Flash & Spark & MOSS \\
\midrule
zh & 337 & 72.4 & 74.8 & 55.8 & 65.9 & 66.5 & 56.4 & 22.6 & 25.8 \\
en & 347 & 75.8 & 75.2 & 62.8 & 66.3 & 65.1 & 53.9 & 51.3 & 25.4 \\
es & 316 & 71.5 & 70.6 & 57.3 & 60.1 & 51.9 & 43.7 & 36.4 & 16.5 \\
fr & 338 & 76.3 & 73.1 & 63.3 & 63.9 & 58.6 & 47.3 & 39.1 & 15.4 \\
ja & 320 & 77.5 & 78.8 & 69.1 & 66.6 & 64.7 & 53.1 & 36.6 & 21.6 \\
ko & 333 & 66.4 & 68.5 & 57.4 & 55.3 & 52.6 & 44.1 & 25.8 & 16.8 \\
id & 324 & 73.5 & 74.4 & 64.5 & 58.3 & 59.3 & 46.0 & 35.5 & 9.6 \\
ru & 308 & 72.1 & 71.4 & 59.7 & 63.0 & 59.4 & 49.7 & 31.8 & 17.9 \\
ar & 313 & 71.2 & 72.5 & 63.9 & 59.4 & 52.7 & 41.9 & 31.6 & 3.5 \\
th & 309 & 72.5 & 72.2 & 58.6 & 57.9 & 50.2 & 43.0 & 31.1 & 12.3 \\
vi & 288 & 76.7 & 78.1 & 61.1 & 63.5 & 51.4 & 42.4 & 26.4 & 4.2 \\
de & 304 & 73.7 & 73.4 & 62.5 & 57.6 & 55.9 & 45.7 & 39.1 & 9.2 \\
it & 304 & 71.1 & 65.1 & 55.6 & 56.2 & 49.7 & 38.2 & 33.2 & 9.9 \\
pt & 306 & 73.2 & 73.5 & 62.7 & 60.1 & 56.9 & 49.3 & 38.2 & 14.7 \\
hi & 290 & 73.4 & 71.7 & 61.4 & 57.6 & 52.8 & 44.1 & 30.7 & 6.9 \\
tr & 301 & 77.4 & 76.7 & 65.4 & 61.8 & 60.5 & 43.5 & 32.6 & 1.7 \\
\bottomrule
\end{tabular*}
\end{table}

\begin{table}[H]
\centering
\caption{Semantic domain accuracy (\%), with 480 questions per domain. Equal cell sizes permit direct averaging over languages.}
\label{tab:domain_results}
\small
\setlength{\tabcolsep}{3pt}
\begin{tabular*}{\textwidth}{@{\extracolsep{\fill}}lrrrrrrrr@{}}
\toprule
Domain & G3.8 & G3.7 & G3.1 & Q-Plus & Doubao & Q-Flash & Spark & MOSS \\
\midrule
Knowledge & 80.0 & 78.5 & 74.6 & 70.2 & 67.5 & 55.6 & 39.6 & 16.0 \\
How-to & 77.1 & 77.7 & 70.0 & 69.4 & 64.4 & 54.8 & 36.2 & 14.8 \\
Narrative & 78.1 & 76.5 & 68.5 & 63.3 & 58.1 & 48.7 & 28.1 & 11.5 \\
Entertainment & 80.8 & 82.5 & 73.1 & 64.2 & 64.4 & 49.0 & 35.8 & 15.8 \\
Lifestyle & 80.8 & 80.6 & 72.1 & 64.6 & 66.7 & 53.5 & 36.2 & 15.8 \\
News & 77.1 & 77.5 & 73.3 & 67.3 & 61.3 & 51.9 & 38.1 & 10.8 \\
Product & 77.7 & 78.8 & 71.9 & 63.3 & 61.2 & 51.0 & 38.5 & 12.5 \\
Belief & 80.8 & 79.8 & 73.5 & 72.1 & 59.4 & 55.0 & 42.3 & 11.5 \\
\bottomrule
\end{tabular*}
\end{table}

\subsection{Evidence by Task}

\begin{table}[H]
\centering
\caption{Semantic accuracy (\%) for all ten models. Denominators are 1024, 1024, 896, 896 in column order.}
\label{tab:semantic_tasks}
\small
\setlength{\tabcolsep}{3pt}
\begin{tabular*}{\textwidth}{@{\extracolsep{\fill}}lrrrr@{}}
\toprule
Model & Factual & Reasoning & Temporal & Long-range \\
\midrule
Gemini 3.8 Flash & 90.33 & 73.44 & 66.07 & 85.60 \\
Gemini 3.7 Flash & 91.41 & 72.85 & 65.40 & 85.38 \\
Gemini 3.1 Pro & 89.75 & 70.21 & 40.96 & 85.38 \\
Qwen3.5-Omni-Plus & 84.28 & 60.35 & 39.96 & 81.03 \\
Doubao seed 2.0 Lite & 82.03 & 60.45 & 29.02 & 77.57 \\
Qwen3.5-Omni-Flash & 72.17 & 43.55 & 27.46 & 65.07 \\
Muse Spark 1.2 & 53.22 & 30.37 & 14.73 & 47.77 \\
\makecell[l]{MOSS-Audio-8B\\Thinking} & 18.65 & 11.33 & 6.70 & 17.30 \\
\makecell[l]{Qwen3-Omni\\30B-A3B-Instruct} & 45.70 & 21.97 & 12.17 & 35.04 \\
\makecell[l]{Audio Flamingo\\Next 8B} & 25.78 & 10.74 & 5.69 & 18.08 \\
\bottomrule
\end{tabular*}
\end{table}

\begin{table}[H]
\centering
\caption{Acoustic accuracy (\%) for all ten models. Denominators are 867, 122, 112, 97 in column order.}
\label{tab:acoustic_tasks}
\small
\setlength{\tabcolsep}{3pt}
\begin{tabular*}{\textwidth}{@{\extracolsep{\fill}}lrrrr@{}}
\toprule
Model & Factual & Reasoning & Temporal & Long-range \\
\midrule
Gemini 3.8 Flash & 48.79 & 64.75 & 71.43 & 82.47 \\
Gemini 3.7 Flash & 46.48 & 69.67 & 74.11 & 82.47 \\
Gemini 3.1 Pro & 17.07 & 57.38 & 28.57 & 71.13 \\
Qwen3.5-Omni-Plus & 35.18 & 63.93 & 45.54 & 72.16 \\
Doubao seed 2.0 Lite & 33.91 & 59.02 & 8.93 & 79.38 \\
Qwen3.5-Omni-Flash & 19.38 & 51.64 & 29.46 & 69.07 \\
Muse Spark 1.2 & 18.69 & 63.11 & 0.89 & 57.73 \\
\makecell[l]{MOSS-Audio-8B\\Thinking} & 9.34 & 40.98 & 7.14 & 18.56 \\
\makecell[l]{Qwen3-Omni\\30B-A3B-Instruct} & 7.73 & 36.89 & 2.68 & 24.74 \\
\makecell[l]{Audio Flamingo\\Next 8B} & 5.88 & 25.41 & 1.79 & 14.43 \\
\bottomrule
\end{tabular*}
\end{table}

\subsection{Headline Accuracy Intervals}

\begin{table}[H]
\centering
\caption{Overall accuracy and Wilson 95\% intervals over the fixed question set.}
\label{tab:headline_intervals}
\small
\setlength{\tabcolsep}{3pt}
\begin{tabular*}{\textwidth}{@{\extracolsep{\fill}}lrrr@{}}
\toprule
Model & Correct / total & Accuracy (\%) & 95\% CI \\
\midrule
Gemini 3.8 Flash & 3,698 / 5,038 & 73.40 & [72.16, 74.60] \\
Gemini 3.7 Flash & 3,684 / 5,038 & 73.12 & [71.88, 74.33] \\
Gemini 3.1 Pro & 3,089 / 5,038 & 61.31 & [59.96, 62.65] \\
Qwen3.5-Omni-Plus & 3,069 / 5,038 & 60.92 & [59.56, 62.26] \\
Qwen3.5-Omni-Flash & 2,345 / 5,038 & 46.55 & [45.17, 47.93] \\
Doubao seed 2.0 Lite & 2,867 / 5,038 & 56.91 & [55.54, 58.27] \\
Muse Spark 1.2 & 1,712 / 5,038 & 33.98 & [32.69, 35.30] \\
\makecell[l]{MOSS-Audio-8B\\Thinking} & 679 / 5,038 & 13.48 & [12.56, 14.45] \\
\makecell[l]{Qwen3-Omni\\30B-A3B-Instruct} & 1,255 / 5,038 & 24.91 & [23.74, 26.12] \\
\makecell[l]{Audio Flamingo\\Next 8B} & 685 / 5,038 & 13.60 & [12.68, 14.57] \\
\bottomrule
\end{tabular*}
\end{table}

\subsection{Residual Variation across Models}
\label{app:language_domain_interaction}
\begin{figure}[H]
    \centering
    \includegraphics[width=\textwidth]{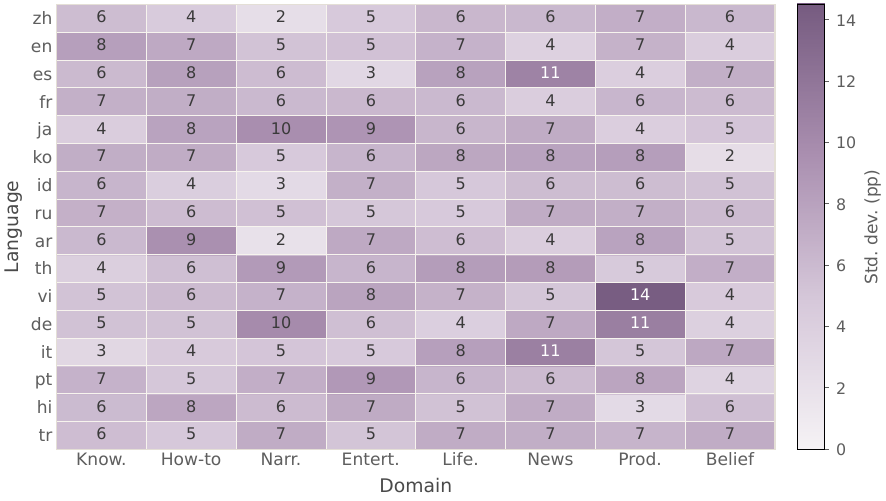}
    \caption{Cross-model standard deviation of Language--Domain additive
    residuals on the balanced semantic track, in percentage points (pp).
    The eight-model diagnostic subset and the residual definition are the same as
    in Figure~\ref{fig:language_domain}(a). Larger values indicate greater
    variation in the interaction residual across models.}
    \label{fig:language_domain_std}
\end{figure}

Figure~\ref{fig:language_domain_std} complements the cross-model mean residuals
in Figure~\ref{fig:language_domain}(a) with their cross-model standard
deviations. The interactions are not uniformly shared across models:
Vietnamese--Product has a residual standard deviation of 14 pp, while
Spanish--News, German--Product, and Italian--News are also highly
model-sensitive (about 11 pp). Thus, multilingual difficulty is
\emph{domain-conditioned}, and the conditioning can itself be model-dependent.

\subsection{Adjusted Evidence--Task and Distance Contrasts}
Table~\ref{tab:regression_full} reports all evidence--task categories and
continuous terms from the identifiable eight-model fit. These estimates use
the specification in Appendix~\ref{app:statistics}; odds ratios and interval
endpoints are reported to three decimals.

\begin{table}[H]
\centering
\caption{Adjusted odds ratios from the eight-model fit. The first seven rows compare against semantic factual; remaining rows represent +1 SD shifts.}
\label{tab:regression_full}
\small
\setlength{\tabcolsep}{3pt}
\begin{tabular*}{\textwidth}{@{\extracolsep{\fill}}lrr@{}}
\toprule
Contrast & OR & 95\% CI \\
\midrule
Acoustic factual & 0.096 & [0.084, 0.109] \\
Acoustic temporal & 0.105 & [0.084, 0.129] \\
Semantic temporal & 0.147 & [0.127, 0.167] \\
Semantic reasoning & 0.332 & [0.291, 0.380] \\
Acoustic reasoning & 0.385 & [0.289, 0.525] \\
Acoustic long-range & 0.587 & [0.433, 0.830] \\
Semantic long-range & 0.834 & [0.723, 0.955] \\
Log duration (+1 SD) & 0.798 & [0.762, 0.832] \\
Relative earliest position (+1 SD) & 0.897 & [0.854, 0.945] \\
Relative evidence span (+1 SD) & 0.961 & [0.912, 1.015] \\
Evidence count (+1 SD) & 1.046 & [0.999, 1.090] \\
\bottomrule
\end{tabular*}
\end{table}
\FloatBarrier
\section{Prompt and Verification Interfaces}
\label{app:prompts}
This appendix describes the inputs and output decisions of the construction
and evaluation calls. These are interface summaries, not verbatim prompt
transcriptions. Programmatic gates are distinguished from model judgments.

\subsection{Construction Calls}
\paragraph{Evidence mining and planning.}
Semantic mining receives timestamped subtitle cues and returns self-contained
facts with supporting times. Planning organizes supported facts into retrieval,
enumeration, multi-slot, multi-hop, and temporal structures. Rebase and packing
constraints are enforced as described in Appendix~\ref{app:construction}.

\paragraph{Question writing and answer derivation.}
The question writer receives the evidence plan and emits the question only,
without a prewritten gold answer. A separate fresh-context call receives the
question and supporting evidence and derives the reference answer. The two
calls use the same model family; independence refers to call context, not to
different model providers.

\paragraph{Acoustic verification.}
The first Gemini pass receives the selected source clip without TimeAudio's
proposal and records perceived non-speech events. The second pass receives the
proposal and blind observations and returns keep, modify, or delete decisions.
Each retained event must be audible; temporal items additionally require
sufficiently precise timing. Localized audibility checks reject uncertain
cited events before final review.

\paragraph{Shortcut checks.}
A question-only solver receives no audio, transcript, or supporting evidence;
its answer is compared against the reference. For acoustic transcript checks,
a solver receives the transcript and applies the task-specific criterion in
Eq.~\ref{eq:acoustic_ablation}. A second verifier confirms whether a proposed
transcript-only answer materially matches the reference before the item is
removed.

\subsection{Inference and Judge Interfaces}
The evaluated model receives canonical audio and the source-language question,
with a fixed language-specific wrapper equivalent to ``Please answer the
following question based on the full audio content.'' It receives no source
transcript or gold evidence. This English wording specifies the meaning of the
wrapper, not the exact text used for every language.

Each judge call receives the question, reference answer, and model's final
answer under one of the six rubrics in Table~\ref{tab:judge_rules}.
Single-fact uses low reasoning effort; the remaining routes use medium effort.
The judge performs semantic comparison, while programmatic checks enforce
required slots and temporal matching. The point/interval tolerance is applied
by code after event--time pairing, rather than delegated to an LLM.
\section{Qualitative Cases, Release, and Limitations}
\label{app:cases}\label{app:ethics}
\subsection{Illustrative Response Cases}
The following cases illustrate response differences; they are not estimates of
error frequencies. They are drawn from the fixed diagnostic cohort.

\paragraph{Correct event, incorrect clock (Turkish).}
A how-to recording states that a stitch count increases from 63 to 69 at 16:06.
Several model responses identify that change but report 15:57, 16:26, or 14:32.
The event is recovered, while its alignment to physical time is incorrect.

\paragraph{Natural event confusion (Portuguese and Italian).}
For a Portuguese factual question about an event near 33:48, the verified event
is a man's laugh. Gemini 3.8 Flash, Gemini 3.7 Flash, Qwen3.5-Omni-Plus, and
Doubao identify laughter; Gemini 3.1 Pro instead reports victory music from
\textit{Final Fantasy}. In an Italian how-to item targeting a short laugh,
responses propose a mouse click, cough, or notification sound. These cases
illustrate local event-grounding errors in natural recordings.

\paragraph{Entity recovery in long-range retrieval (Korean).}
The recording gives a past sentence of ``2 years 6 months with 4 years
suspended.'' Several models recover the pair, while one changes the prison
term to 1 year 6 months and another selects a separate life-imprisonment
conviction mentioned in the same recording. The item distinguishes precise
entity recovery from grounding to a nearby competing event.

\subsection{Release Scope and Provenance}
The benchmark artifacts comprise questions, reference answers, localized
evidence, task metadata, and evaluation metadata. Source provenance supports
traceability. Original recordings were collected from publicly accessible
online media; copyright remains with the respective rights holders.
Redistribution of source audio is subject to source-specific permissions.

\subsection{Limitations}
MuLA-Bench controls semantic question allocation across languages and domains,
but it does not hold recording content or speakers identical across languages.
Acoustic availability and task mixtures remain natural and uneven; the
acoustic--semantic gaps are descriptive, not matched interventions. Long-range
semantic questions include late single facts as well as widely separated
evidence, so strong retrieval does not establish general long-range reasoning.
Local context limits and response failures remain part of the end-to-end score.
The 117-response judge audit is a sampled agreement check, and final dataset
review uses one expert per language rather than independent duplicate
annotation. Pooled diagnostic claims concern the fixed eight-model cohort.

\end{document}